\documentclass[twocolumn]{aastex63}
\usepackage{graphicx}
\usepackage{txfonts}
\usepackage{setspace} 
\usepackage{booktabs}
\usepackage[inline, shortlabels]{enumitem}

\def\ie{{{\rm i.e.,} }}
\def\be{\begin{equation}}
\def\bea{\begin{eqnarray}}
\def\eea{\end{eqnarray}}
\def\ee{\end{equation}}

\def\g0{\Gamma_{0}}

\def\Q{\cal{Q}}

\def\Sta{\rm{St}_1}
\def\Stb{\rm{St}_2}
\def\vcrit{v_{\rm crit}}
\def\vrel{v_{\rm rel}}
\def\ropeb{\rho_{\rm p}}
\def\PB{{\cal{P}_{\rm B}} }
\def\lmax{{l_{\rm max}}}
\def\xmax{{x_{\rm max}}}

\newcommand{\tcoll}{t_{\text{coll}}}
\newcommand{\tdrag}{t_{\text{drag}}}

\newcommand{\Lagr}{\mathcal{L}}
\newcommand{\rvec}{\mathbf{r'}}
\newcommand{\Rvec}{\mathbf{R}}

\newcommand{\muval}{\mu}
\newcommand{\vvec}{\dot{\rvec}}
\newcommand{\xval}{x}

\shorttitle{Planetesimal formation through screening interaction}
\graphicspath{{./}{figures/}}

\begin{document}
\setstretch{1.0}
\title{Gas Pressure Driven Screening Forces and Pebble Aggregation:\\ A Pathway for Growth in Planet Formation}

\correspondingauthor{Mukesh Kumar Vyas} 
\email{mukeshkvys@gmail.com}


\author{Mukesh Kumar Vyas}
\affiliation{Bar Ilan University, \\ Ramat Gan, Israel,
 5290002}
\begin{abstract}
The formation of planetesimals from cm-sized pebbles in protoplanetary disks faces significant barriers, including fragmentation and radial drift. We identify a previously unaccounted screening force, arising from mutual shielding of thermal gas particles between pebbles when their separation falls below the gas mean free path. This force facilitates pebble binding, overcoming key growth barriers under turbulent disk conditions. Unlike conventional mechanisms, screening forces operate independently of surface adhesion and complement streaming instability and pressure traps by enhancing aggregation in high-density regions. Our analysis predicts that screening interactions are most effective in the {middle disk regions ($ \sim 0.3$ to few AU),} consistent with ALMA observations (e.g., TW Hya) of enhanced dust concentrations. {Furthermore, we find that screening-induced pebble growth from centimeter to kilometer scales can occur on timescales significantly shorter than the disk lifetime ($\sim 10^5$ years). Importantly, this growth naturally terminates when particles smaller than the local gas mean free path are depleted, thereby avoiding runaway accretion.}
Beyond planetary science, the screening forces have {potential} implications for high-energy astrophysics, dusty plasmas, confined particle suspensions and other relevant areas, suggesting a broader fundamental significance.
\end{abstract}
\keywords{}

\section{Introduction}
\label{sec_intro}

The formation of planetesimals, kilometer-sized bodies that serve as the fundamental building blocks of planets, remains one of the most enigmatic stages of planet formation \citep{1916JRASC..10..473C, 1993ARA&A..31..129L, 2007Natur.448.1022J, 2009ApJ...704L..75J, 2023ASPC..534..863W}. The core accretion model was developed to explain the formation of giant planets within the framework of the planetesimal hypothesis \citep{1980PThPh..64..544M, 1996Icar..124...62P, 1987Icar...69..249L}. Observations of protoplanetary disks at millimeter and centimeter wavelengths indicate that micrometer-sized dust grains consistently grow to pebble-sized particles around nearly all young stars \citep{2003A&A...403..323T, 2005ApJ...626L.109W}. 

However, the transition from pebbles to planetesimals faces multiple theoretical challenges. The most significant barriers to planetesimal formation include insufficient pebble densities to trigger gravitational collapse \citep{2008ARA&A..46...21B}, rapid inward drift of pebbles due to gas drag \citep{1977MNRAS.180...57W, 2010MNRAS.404..475J}, and collisional fragmentation \citep{2017AREPS..45..359J, 2021NatRP...3..405W} at high impact velocities \citep{2011A&A...525A..11B}. At lower velocities, the bouncing barrier prevents pebbles from sticking upon collision \citep{2009PhRvL.103u5502K, 2016ApJ...827..110K, 2010A&A...513A..57Z}. Even if the pebbles survive these barriers, they experience rapid radial drift, spiraling into the central star before forming planetesimals. These obstacles severely restrict dust growth, necessitating additional mechanisms such as pressure traps, turbulence, or ice mantles to facilitate planet formation \citep{1977MNRAS.180...57W, 2010MNRAS.404..475J}. To bind the pebbles together, contact forces such as van der Waals forces have been criticized as a potential binding mechanism to induce stable pebble growth beyond micron-sized grains \citep{2020apfs.book.....A}. 

Among the proposed mechanisms, streaming instability provides a promising route for planetesimal formation by inducing strong local concentrations of pebbles via dust-gas interactions \citep{2005ApJ...620..459Y, 2007Natur.448.1022J, 2023ApJ...959L..15Z}. Numerical simulations show that under favorable conditions, such as enhanced dust-to-gas ratios and low turbulence, pebbles cluster and collapse into planetesimals \citep{2007ApJ...662..627J, 2009ApJ...704L..75J}. However, the efficiency of streaming instability depends on specific disk conditions and particle size distributions, making its universal applicability uncertain. Additionally, pressure bumps and vortices, which can serve as pebble traps, are often transient and highly dependent on local disk structures \citep{1995A&A...295L...1B,2013ApJ...775...17L}. {Other in situ formation models, such as Inside-Out Planet Formation, where cm-sized pebbles are trapped in rings at the dead-zone inner boundary, forming planets sequentially \citep{2014ApJ...780...53C, 2015ApJ...798L..32C, 2016ApJ...816...19H}, predicting efficient planet formation in low-viscosity disks with viscosity parameter $\alpha \sim 10^{-4}$ \citep{2018ApJ...857...20H}.} Despite the effectiveness of these mechanisms in certain environments, a more general process is needed to promote pebble growth as the disks are found to be capable of producing the planets very efficiently \citep{2020apfs.book.....A}.

In this study, we identify \textit{\bf screening force}, a novel mechanism that naturally arises from scattering induced anisotropic gas pressure between interacting pebbles. When the separation between pebbles falls below the gas mean free path, mutual shielding creates a pressure imbalance that results in a net attractive force. This effect enhances pebble binding, enabling growth in conditions where traditional sticking mechanisms fail. By modeling the strength and spatial dependence of the screening force, we demonstrate its ability to overcome growth barriers even in turbulent protoplanetary disks. Furthermore, this mechanism is more effective in the presence of streaming instability and other density-enhancing instabilities, complementing existing formation pathways.

Our work connects the theoretical framework of the screening force to observable signatures of planetesimal formation, particularly the substructures in protoplanetary disks identified by high-resolution surveys such as the Atacama Large Millimeter/submillimeter Array (ALMA) and the Disk Substructures at High Angular Resolution Project (DSHARP). We predict enhanced binding probabilities in the {middle regions of disks (\(\sim 0.3$ to a few $\, \mathrm{AU}\)), where} gas density and temperature conditions are optimal for screening forces to operate. This study not only addresses a fundamental challenge in planetary science but also provides a framework for interpreting the observed grain size distributions and dust substructures in protoplanetary disks \citep{2016ApJ...820L..40A, 2018ApJ...869...17L}. 

Beyond implications for planetesimal formation, the screening interaction represents a general physical process that extends beyond planetary science. The underlying mechanism (anisotropic pressure shielding leading to attractive forces) can be relevant in a variety of fields, including condensed matter physics, soft matter interactions, and biological systems where small-scale forces govern aggregation dynamics. Recognizing this broader applicability strengthens the interdisciplinary impact of our findings.

\section{Theory of screening forces}
\label{sec_theory}
\begin{figure}[h!]
    \centering
     \includegraphics[width=8 cm]{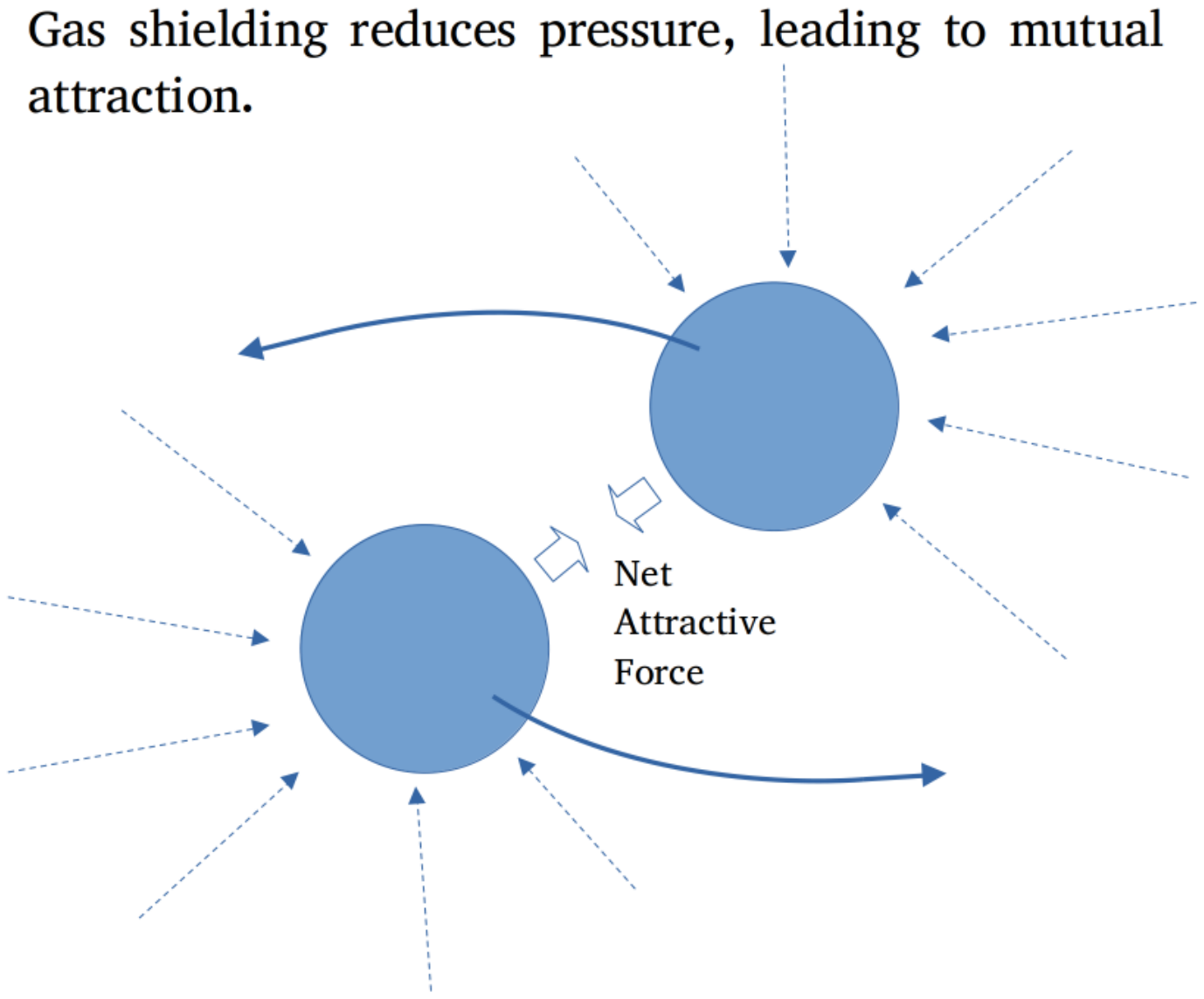}
    \includegraphics[width=8 cm]{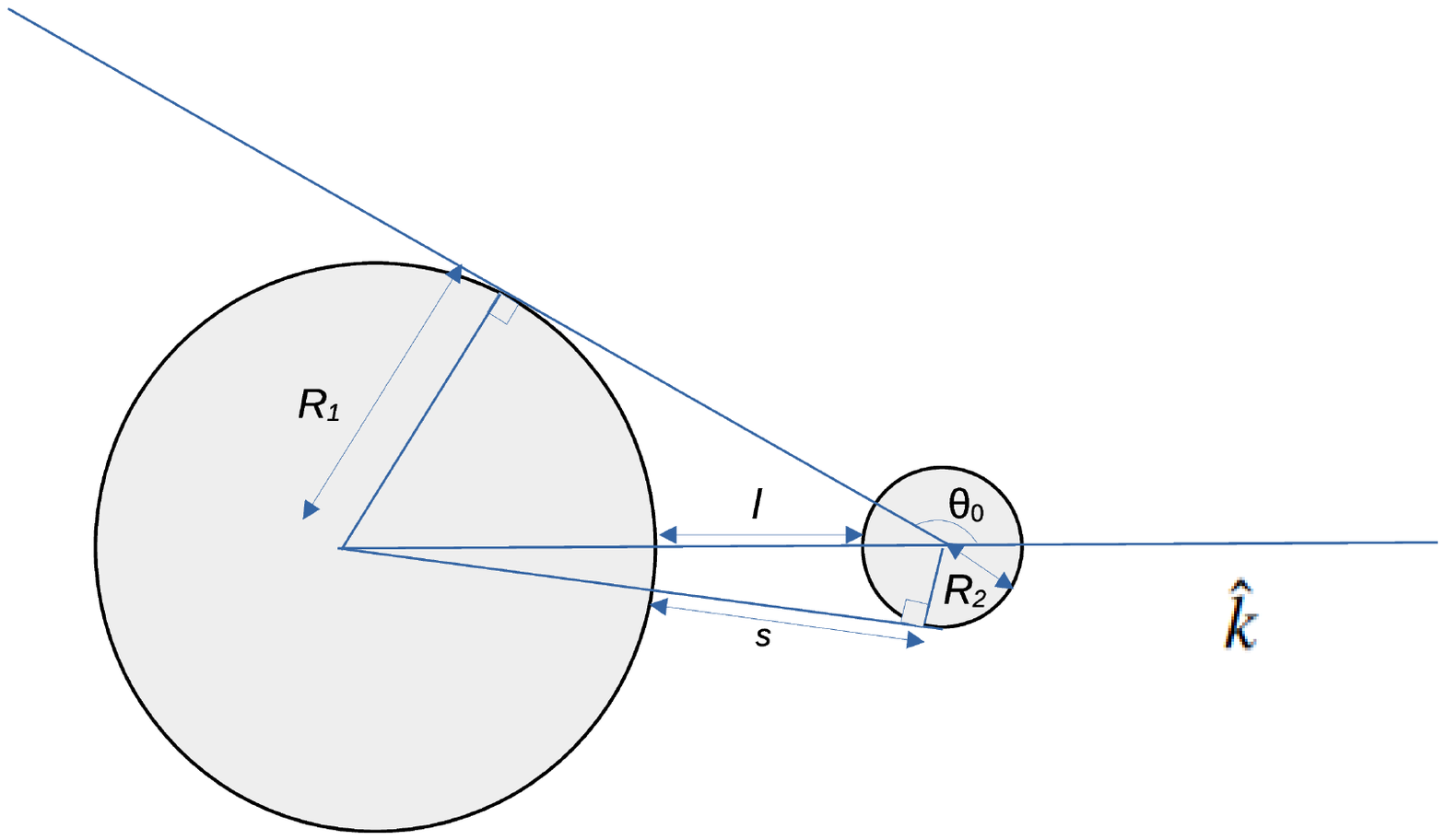}
    \caption{Top: Visual representation of screening forces as a result if screened thermal particles from opposite sides. The screened region between the spheres causes an attractive force due to the imbalance in gas pressure. Bottom: Illustration of screening force between two spheres in a gas. {A nearby larger sphere with radius $R_1$ blocks the angular region $\theta_0 \leq \theta \leq \pi$ on smaller sphere with radius $R_2$, resulting in a net force $\mathbf{F}$ due to anisotropic pressure acting over $0 \leq \theta \leq \theta_0$. The force is effective for condition $s\le\lambda$.}
    }
    \label{lab_screening_force}
\end{figure}

The screening force arises when two bodies in a gas medium experience a net attraction due to a local pressure imbalance. When their mutual separation becomes comparable to the mean free path of gas molecules, each body blocks the gas flux reaching the other’s facing side, reducing the pressure between them and creating a net force that draws them together (Figure \ref{lab_screening_force}, top panel). This effect may play a crucial role in the slow aggregation of particles in gas-rich environments, including protoplanetary disks.
\subsection{Computation of screening force}
To estimate the net attractive force between two pebbles, consider a gas with number density \( n \), temperature \( T \) and the isotropic pressure $p$.
Two spherical particles (or pebbles) with radii \( R_1 \) and \( R_2 \,(\le R_1) \), and masses \( M_1 \) and \( M_2 \), are separated by a surface-to-surface distance \( l \) (Figure \ref{lab_screening_force}, bottom panel). We define a spherical coordinate system with origin being at the center of Sphere 2. The screening reduces the gas pressure on their adjacent sides if $s\le \lambda$ where $\lambda$ is the mean free path of the gas particles {and $s=\sqrt{x^2-R_2^2}-R_1$ with} $x=R_1+R_2+l$. {The condition \( s < \lambda \) ensures that the second particle lies within the collisionless shadow cast by the first, preserving the anisotropy in the background gas momentum flux. If \( s > \lambda \), gas particles scatter and re-isotropize between the two, eliminating the screening effect. The quantity \( s \) is defined as the distance from the surface of the larger particle to the tangent point on the smaller one, ensuring that the entire smaller sphere lies within the shadow cone. This guarantees full blockage of the background momentum flux over the solid angle responsible for the screening force.} 
This pressure difference creates a net attractive force between the bodies that can be quantified as follows. 
The net force on an isolated sphere in an isotropic gas is zero: \(\mathbf{F} = p \oint \mathrm{d}\mathbf{A} = 0 \), where \( \mathrm{d}\mathbf{A} \) is the vector area element on the sphere {with bounded surface area ${A}$}. The presence of a nearby larger sphere blocks the {angular region $\theta_0$ to $\pi$} so the net force ${F}$ due to anisotropy in pressure now acts over the angular {space \( 0 \leq \theta \leq \theta_0 \) }(Figure \ref{lab_screening_force}, bottom panel),
\be 
{F} = p \int_0^{2\pi} \int_0^{\theta_0}R_2^2\hat{r} \sin \theta d\theta d\phi \\ \\  
\ee 
Where $\hat{r}=\sin \theta \cos \phi \hat{i}+\sin \theta \sin \phi \hat{j}+\cos \theta \hat{k}$ is the unit vector pointing outward the surface of the sphere 2, the integration returns the force on sphere 2 due to the presence of bigger sphere 1,
\be 
{F} = -\pi p  R_2^2 \sin^2{\theta_0} \, \hat{k} = \frac{-\pi n k_{B} T R_1^2 R_2^2}{x^2}\,  \hat{k}  ~~~~ {\rm (For~~ }s \leq \lambda)
\label{eq_F}
\ee 
 Here $k_B$ is Boltzmann's constant. As this is a mutual attractive force between the spheres, the same force acts on sphere 1 due to sphere 2.
The force vanishes for separations larger than the gas mean free path ($s>\lambda$) \footnote{{However, this theoretical cutoff is not abrupt. In practice, the force decays smoothly at $s\sim\lambda$ , becoming negligible for ${F}=0$ at $s>>\lambda$.}}.
When the bodies move randomly and interact, for $l<0$, the pebbles collide, potentially leading to fragmentation.
The force increases with particle size (\( F \propto R_1^2 R_2^2 \)), scales linearly with gas density and temperature, and follows an inverse-square dependence on distance. The inverse-square nature arises because the blocked flux depends on the solid angle subtended by one sphere on the other, which scales inversely with the square of the separation. The force vanishes for point size particles. The maximum force is at $l\rightarrow 0$ when the two particles are in contact. Further, when a comparatively very small size particle comes in the vicinity of a large particle $R_1>>R_2, \, {\rm and}\, \, l \rightarrow 0$, we have $\theta_0\rightarrow \pi/2$ and hence $F \propto -pA_2$ where $A_2$ is the surface area of the smaller pebble. {Unlike typical attractive forces such as van der Waals or Coulomb interactions, which are weakened by thermal agitation, the screening force uniquely strengthens with increasing background gas temperature, as it arises from thermal momentum flux that scales with \( k_B T \).
}

As the screening forces vanish at distances larger than the mean free path, despite the identical sizes of systems around us ($\sim $cm), the force is not perceptible in daily life as the mean free path of air molecules is comparatively much smaller ($10^{-7}$ cm). In the protoplanetary disks, the mean free path between the gas particles is large enough ($\sim$ cm), making the screening force an effective candidate to imply an attractive force between pebbles.

The central idea of screening forces is similar to what Nicolas Fatio de Duillier and Georges-Louis Le Sage \citep{le1782nouveaux} described as kinetic theory of gravity, which now is obsolete \citep{le2022sage}. However, the phenomenologically similar mechanism exists for particles in other situations such as interactions of particles with light in interstellar medium \citep{1941ApJ....94..232S}, particles in dusty plasma \citep{1996PlPhR..22..585I} and some cosmological implications in the early universe \citep{1949RvMP...21..367G}. {The current work identifies a novel version of the mechanism in a gaseous medium effective at mutual distances shorter than the gas mean free paths between interacting particles.
}
\begin{figure}[h!]
    \centering
    \includegraphics[width=7 cm, trim = 0cm 7cm 3cm 2cm, clip]{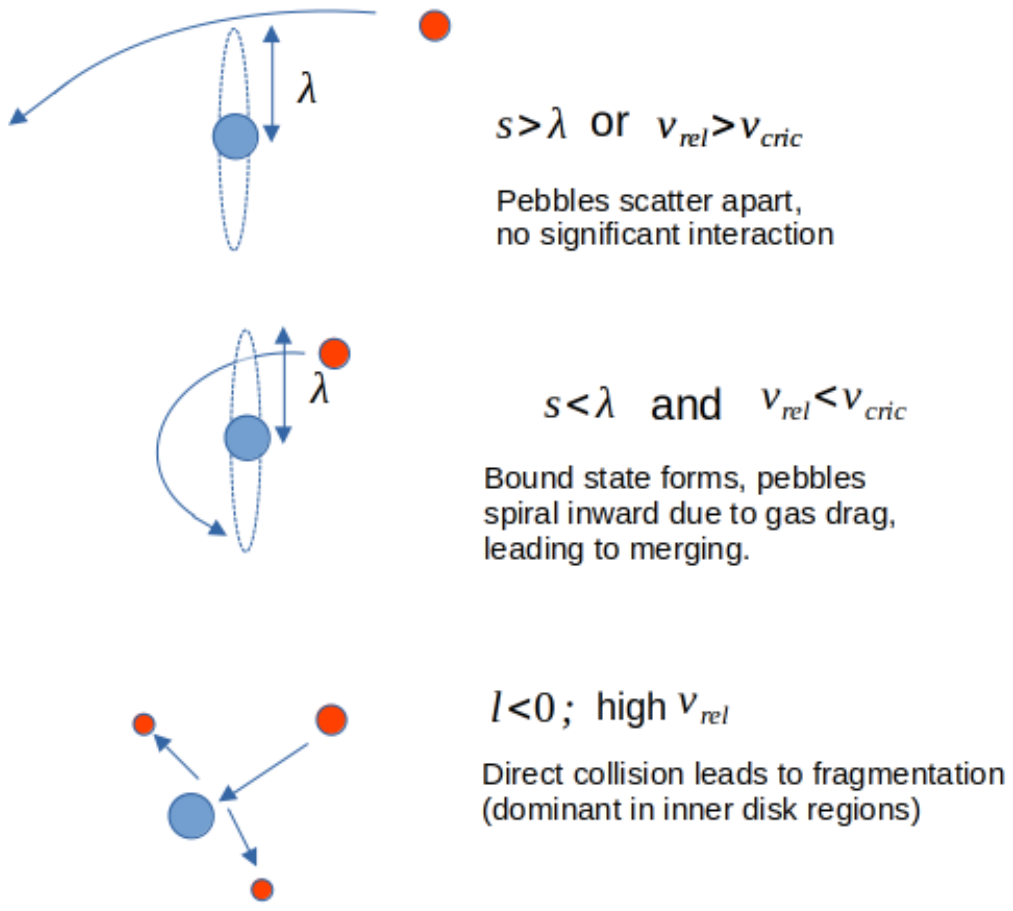}
    \caption{Geometry of scattering between two pebbles. \textit{Top}: no bound state if $s>\lambda$ or $\vrel>\vcrit$, \textit{middle}: bound state forms for opposite case followed by in-spiral due to gas drag leading to merging, \textit{bottom}: fragmentation due to direct collision if $l<0$, dominant process in the inner disc region where binding probability drops.}
    \label{lab_geom_of_scattering}
\end{figure}

\subsection{Binding energy of pebbles via screening forces and critical velocities}
\label{sec_theory_2}
We now analyze the effectiveness of screening forces in binding two pebbles during encounters within $s\leq \lambda$. 
While the previous section considers the force between static bodies, in protoplanetary disks the pebbles experience random motion due to turbulence. This random motion works against any attractive interaction. To determine the maximum relative velocity that allows binding (the critical velocity \( v_{\rm crit} \)), we compare the pebbles' kinetic energy (KE) with the work required to overcome the attractive potential energy (PE). The condition for binding is determined by ${\rm KE} + {\rm PE} < 0,$ or equivalently:

\be 
\frac{1}{2} \mu v_{\rm rel}^2 < 
-\int_{\xmax}^{x} F(\tilde{x}) \, d\tilde{x}= -\pi n k_B T R_1^2 R_2^2\left(\frac{1}{x}-\frac{1}{\xmax} \right)
\label{eq_PE}
\ee 

Here \( \mu = {M_1M_2}/({M_1+M_2}) \) is the reduced mass of the pebbles, \( v_{\rm rel} \) is relative velocity between them,  {$\xmax=\sqrt{(\lambda+R_1)^2+R_2^2}$ is the maximum distance between the centres of the spheres for which the screening force is active.}
For intrinsic density of pebbles being $\ropeb$, the kinetic energy of the two pebbles is:
\[
{\rm KE} = \frac{1}{2} \mu v_{\rm rel}^2 = \frac{2\pi R_1^3R_2^3 \ropeb v_{\rm rel}^2}{3(R_1^3+R_2^3)},
\]
{Equating \( {\rm KE} = {\rm PE} \), we define a critical velocity $v_{\rm crit}$  (\ie $v_{\rm rel} = v_{\rm crit}$),}


\be
v_{\rm crit} = \sqrt{\frac{3 n k_B T (R_1^3+R_2^3) }{2R_1R_2\ropeb}\left(\frac{1}{x}-\frac{1}{\xmax} \right)}.
\ee
Using defined expressions of the energies above, one can derive the screening force from Lagragian formalism considering it to be a conservative force\footnote{{Using relative coordinates $\rvec = \rvec_1 - \rvec_2$ with center-to-center distance $\xval = |\rvec_1 - \rvec_2|$ and center-of-mass coordinates $\Rvec = (M_1 \rvec_1 + M_2 \rvec_2)/(M_1 + M_2)$, the Lagrangian for the relative motion is (For $s<\lambda$),
\begin{equation}
\Lagr(\rvec, \vvec) = \frac{1}{2} \muval \vvec^2 -
\frac{\pi n k_B T R_1^2 R_2^2}{|\rvec|}.
\end{equation}
Applying Euler-Lagrange equation, one recovers conservative form of the screening force law:
\begin{equation}
\muval \ddot{\rvec} =
-\frac{\pi n k_B T R_1^2 R_2^2}{|\rvec|^2} \hat{r} 
\end{equation}
consistent with Equation \ref{eq_F}.
}}.
{For binding to occur, \( {\rm KE} < {\rm PE} \) and hence the required condition for binding on velocity of the pebble is $v_{\rm rel} < v_{\rm crit}$}, When a small particle passes near a much larger one (\( R_1 \gg R_2 + l \)) without direct collision {and considering $\xmax>>x$}, the critical velocity reduces to:
\be 
v_{\rm crit} = \sqrt{\frac{3nk_BTR_1}{2\ropeb R_2}}.
\label{eq_vcrit}
\ee 
In a turbulent gas in the protoplanetary disk, the relative speed between two pebbles is expressed in terms of the Stokes parameters $\Sta$ and $\Stb$ \citep{2007A&A...466..413O}. 

\be 
\vrel = V_g\sqrt{St_1\left[ 2y_a - (1 + \epsilon) + \frac{2}{1+\epsilon} \left( \frac{1}{1+y_a} + \frac{\epsilon^3}{y_a + \epsilon} \right) \right]}
\label{eq_vrel}
\ee 
Here $y_a = 1.6$, $\epsilon = \Stb/\Sta$, $V_g(=\sqrt{\alpha} c_s)$ is the turbulent gas speed in the disk for turbulence parameter $\alpha$ and sound speed \( c_s\) \cite[$= \sqrt{{\gamma k_B T}/{\mu_m m_p}}$][]{2020apfs.book.....A} , $\gamma$ is adiabatic index of the gas (we consider $\gamma =5/3$ in this study), \(\mu_m = 2.34\) is mean molecular weight, and \(m_p\) is proton mass. The Stokes number (\({\rm St}\)) is defined as the ratio of the particle's stopping time (\(t_s\)) to the orbital timescale (\(1/\Omega_K\)), where \(\Omega_K (= \sqrt{{GM_\star}/{r^3}}\)) is the Keplerian angular velocity \citep{2007A&A...466..413O} with $G$ being the universal constant of gravity.
Thus ${\rm St} = \Omega_K t_s,$ depends on the 
stellar mass \(M_\star\) and the radial distance \(r\) from the star. The stopping time \(t_s\) represents the timescale over which a particle's velocity relative to the gas is damped due to drag.
In the Epstein regime, applicable when the particle size \(R\) is smaller than the mean free path of the gas (\(\lambda\)), the stopping time can be described as \(t_s = {\ropeb R}/{\rho c_s}\). Here \(\rho\) is the gas density, and \(c_s\) is the local sound speed. In the Stokes regime, where \(R > \lambda\), the stopping time is \(t_s = {4 \ropeb R^2}/{9 \rho c_s \lambda}\) \citep{2007Icar..192..588Y, 2007A&A...466..413O,2016SSRv..205...41B}. 
{As the screening interaction is general and effective on all scales of the particle sizes until the distance between the pebbles surpass the mean free path of the surrounding gas, it accounts for particles ranging from Epstein regime to Stokes regime. Further, when a single particle of small size (epstein regime) grows to large size through successive pebble coagulation  (section \ref{sec_growth_time}), it transits to Stocks regime applicable to large size pebbles.}

\subsection{Effect of gas drag and pebble merging}
If the probability of binding is significant (section \ref{sec_binding_probability_1}), once two pebbles form a bound state, their relative kinetic energy is dissipated through gas drag, leading to their eventual merging into a single aggregate. The drag force acting on a pebble {is given as \citep{2011ApJ...733...56P}},

\[
F_{\rm drag} = \frac{1}{2} C_D \rho A v_{\rm rel}^2,
\]
where \( C_D \sim 1 \) is the drag coefficient, and \( A = \pi R^2 \) is the cross-sectional area of the pebble with radius $R$. 
The timescale for kinetic energy dissipation, \( t_{\rm drag} \), is then:

\be 
t_{\rm drag} \sim \frac{\mu v_{\rm rel}}{F_{\rm drag}}, 
\label{eq_tdrag}
\ee
{In case $R_1>>R_2$, the smaller particle experiences drag and gradually merges with the larger one as its relative velocity dissipates and $\mu$ equals the mass of the smaller pebble and $R=R_2$}. 
This ensures that bound pairs efficiently lose energy and merge, initiating a cascade of pebble coagulation that leads to the formation of larger aggregates and ultimately planetesimals. While additional weak forces, such as van der Waals interactions, may assist in the final stages of coalescence, they are not strictly necessary. The screening force alone provides the dominant binding interaction, with its strength reaching a maximum at \( l = 0 \), ensuring robust merging even in the absence of significant surface adhesion.

Figure \ref{lab_geom_of_scattering} summarizes the physical mechanisms governing pebble interactions in a gas-rich environment. The top panel depicts a scattering event where the pebbles escape without forming a bound state. This occurs when the separation exceeds the gas mean free path ($s > \lambda$) or when their relative velocity is greater than the critical binding velocity ($v_{{\rm rel}} > v_{{\rm crit}}$). The middle panel shows a successful binding scenario, where for sufficiently low velocities ($v_{{\rm rel}} < v_{{\rm crit}}$) and a minimal separation within the mean free path ($s \le \lambda$), the pebbles form a bound state. Subsequent energy dissipation via gas drag leads to a gentle merging, forming a larger aggregate. In contrast, the bottom panel shows fragmentation, which occurs when pebbles undergo a direct collision ($l < 0$). Fragmentation is more prevalent in high density inner disk regions where the mean free path is small, increasing the likelihood of collisions over binding. In Sections~\ref{sec_binding_probability}-\ref{sec_binding_probability_1}, we will explore a quantitative estimates of binding, scattering, and fragmentation probabilities under varying disk {parameters discussed below}.

\subsection{Disk parameters}
The screening force is characterized by temperature and density in the accretion disc. 
The temperature in a protoplanetary disk is modeled as a power-law function of the radial distance \( r \) from the central star,
\[
T(r) = T_0 \left(\frac{r}{1 \, \text{AU}}\right)^{-q_T},
\]
where \( T_0 \) is the temperature at \( r = 1 \, \text{AU} \) (typically \( \sim 200{-}300 \, \text{K} \)), and \( q_T \) is the temperature power-law index (commonly \( q_T \sim 0.5{-}0.75 \)). In this work, we adopt \( q_T = 0.5 \).
The gas density in a protoplanetary disk is modeled as a vertically stratified profile with vertical distance $z$ from its midplane \citep{1973A&A....24..337S, 2008ARA&A..46...21B}, 
\[
\rho(r, z) = \rho_0(r) \exp\left(-\frac{z^2}{2 H^2}\right),
\]
where \( \rho_0(r) \) is the midplane density and $H = {c_s}/{\Omega_k}$ is the gas scale height  \citep{2017AREPS..45..359J}. 
The midplane gas density can be given as,
\be 
\rho_0 = \frac{\Sigma_g}{\sqrt{2\pi}H}
\ee 
With surface density $\Sigma_g$ is defined as \citep{2017AREPS..45..359J},
\be 
\Sigma_g = \Sigma_0 f_g \left(\frac{r}{1 \, {\rm AU}}\right)^{-1} {\rm ~ gcm}^{-2}
\label{eq_sigma_g}
\ee 
Here $\Sigma_0$ is surface density assumed at 1 AU for considering $f_g \sim 1$ for a young protoplanetary disc. 
In this paper, we take the gas density as the midplane value, i.e., \( \rho = \rho_0 \).
Subsequently,{ we have $n=\rho/\mu_m m_p$.}
The mean free path of gas particles in terms of the gas {density is defined as $\lambda (r) = {1}/{n \sigma_g}$},
here \(\sigma_g ( = 2 \times 10^{-15} $cm $^2)\) is the collision cross-section for molecular hydrogen \citep{1970mtnu.book.....C, 2020apfs.book.....A}. 
The size distribution of pebbles in protoplanetary disks is commonly described by a power-law function of the form $
N_p(R) \propto a^{-q_N},$
where \( N_p(R) \) is the number density of particles with size \( R \), and \( q_N \) is the power-law index, varies between 3 and 3.7 \citep{1969JGR....74.2531D, 2012A&A...539A.148B} supported by observations such as those from ALMA \citep{2012ApJ...744..162A, 2016ApJ...821L..16C}. In this paper, we take $q_N = 3$.

\subsection{Probability of scattering}
\label{sec_binding_probability}
The probability of pebble to pebble collisions in a protoplanetary disk is critical for understanding the growth of planetesimals. It depends on the local number density of pebbles, their relative velocities, and the collision cross-section. The collision rate ($\Gamma$), in general, is expressed as \citep{2020apfs.book.....A}:

\begin{equation}
    \Gamma = n_{\rm peb} \sigma_{\rm coll} v_{\rm rel},
\end{equation}

where $n_{\rm peb}$ is the number density of pebbles. {The upper limit of $l$ for which the screening force exists is 
$\lmax = \sqrt{(\lambda+R_1)^2+R_2^2}-(R_1+R_2)$.} Hence,  $\sigma_{\rm coll}[=\pi \xmax^2]$ is the collision cross-section within impact parameter $\xmax$. The probability of a collision within a timescale $t$ is then,
\begin{equation}
    P_{\rm sca} = 1 - e^{-\Gamma t}.
\label{eq_psca}
\end{equation}

The number density of pebbles for dust-to-gas ratio $Z$ is \citep{2015A&A...582A.112B},
\begin{equation}
    n_{\rm peb} = \frac{Q_s Z \rho}{M_2},
\end{equation}
Here $Q_s$ is introduced density enhancement factor that accounts for increased density of pebbles due to streaming instability or the pressure enhanced regions such as followed by Rossby wave instabilities. In this case, $Q_s>1$ while  $Q_s = 1$ for normal disc conditions without such instabilities.
{We consider $Z=0.01$ to} be the standard dust to gas ratio.
The timescale $t$ in Equation \ref{eq_psca} can be taken to be the rotation period of the disk $t=2\pi/\Omega_K$. At $r=1$AU, this timescale is around $1$ year for a disk around a solar mass star. However, as shown by \cite{1977MNRAS.180...57W}, a $1$cm particle typically has around $10^4$ years before it spirals inside the star \citep[Also see][]{2020apfs.book.....A}.
The rate of scattering for the particle is $1/\Gamma$. For $t>>1/\Gamma$, $P_{\rm sca} \rightarrow 1$. 
\subsection{Binding Probability of pebbles with random velocities}
\label{sec_binding_probability_1}
When a pebble scatters with another with $s \le \lambda$, it forms a bound pair if the relative speed between the two pebbles ($v$) is below $\vcrit$. 
For a Maxwellian distribution of relative velocities of pebbles in a turbulent disk with average velocity $\vrel$, the Maxwellian probability density function (PDF) for velocities $v$ is,
\be 
f(v) = \sqrt{\frac{2}{\pi}} \frac{v^2}{\sigma^3} e^{-v^2 / 2\sigma^2},
\ee
where \( \sigma (= \vrel/\sqrt{3})\) is the velocity dispersion between the two pebbles. This means that if there are several copies of two interacting particles with given fixed sizes, the average velocity dispersion is the relative speed defined by Equation \ref{eq_vrel}.

For this Maxwellian distribution of pebbles, the fraction of pebble pairs with \( v \le  v_{\rm crit} \) is given by the cumulative distribution function (CDF):
\begin{equation}
P(v < v_{\rm crit}) = \int_0^{v_{\rm crit}} f(v) \, dv = \text{erf}\left( \frac{v_{\rm crit}}{\sqrt{2} \sigma} \right) - \sqrt{\frac{2}{\pi}}\frac{v_{\rm crit}}{\sigma} e^{- \frac{v_{\rm crit}^2}{2 \sigma^2}}.
\label{eq_prob_error_func}
\end{equation}

Here, \( \mathrm{erf}(\cdot) \) denotes the error function. Equation \ref{eq_prob_error_func} shows the probability for a pebble pair to form a bound system if there is a scattering between them for $0<s\le\lambda$. However, for $l<0$, there would be a direct collision between the pebbles and may lead to {either fragmentation or bouncing off the pebble's surface without coagulation}. Hence, for the scattering within $l<\lmax$, the probability of coalesce is only for scattering within $0<l<\lmax$. The total probability of scattering within this region is the ratio of respective cross sections of both processes, 
\bea 
P_2 = \frac{\pi (\lmax+R_1+R_2)^2-\pi (R_1+R_2)^2}{\pi (\lmax+R_1+R_2)^2} \nonumber \\ 
= 1-\frac{(R_1+R_2)^2}{(\lmax+R_1+R_2)^2}.
\label{eq_P_collase}
\eea

Hence from a collection of pebbles at a given location in the disk, the probability of two pebbles of given sizes will to form a bound state ($\PB$) is given by a product of probabilities in Equations  \ref{eq_psca}, \ref{eq_prob_error_func} and \ref{eq_P_collase},
\[ 
\PB = P_2 \times P(v < v_{\rm crit}) \times P_{\rm sca}
\] 
\be 
= \left[1-\frac{(R_1+R_2)^2}{(\lmax+R_1+R_2)^2}\right] \, {\rm erf}\left( \frac{v_{\rm crit}}{\sqrt{2} \vrel} \right)\, (1 - e^{-\Gamma t})
\label{eq_PB}
\ee 
\\ \\

\begin{figure*}[t!]
    \centering
    \includegraphics[width=10 cm, trim = 1cm 5cm 0cm 2cm, clip]{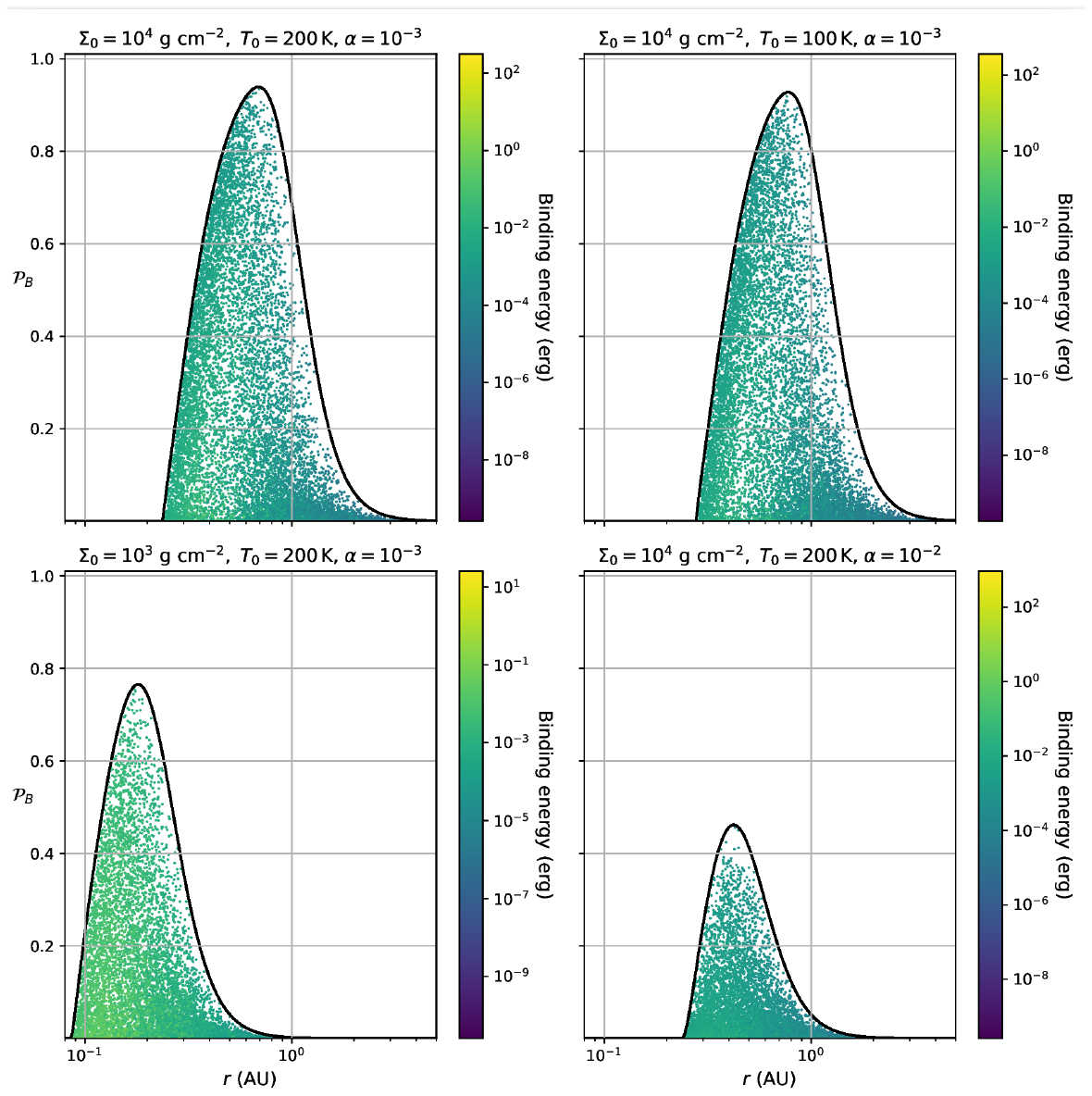}
    \caption{Binding probability of pebble pairs to form a bound state upon scattering in the protoplanetary disk within dynamical timescale of the disk rotation $t$ at given $r$ {For parameters - top left panel: $\Sigma_0 = 10^4$ gcm$^{-2}$, $T_0=200$ K, and $\alpha = 10^{-3}$; top right panel: $\Sigma_0 = 10^4$ gcm$^{-2}$, $T_0=100$ K, $\alpha = 10^{-3}$; bottom left panel: $\Sigma_0 = 10^3$ gcm$^{-2}$, $T_0=200$ K, $\alpha = 10^{-3}$ and bottom right panel $\Sigma_0 = 10^4$ gcm$^{-2}$, $T_0=200$ K, and $\alpha = 10^{-2}$}. The color coding shows the binding energy of pebbles when they merge. {The solid line at the top of each panel denotes the binding probability for the smallest-sized pebble pair in our sample, with $R_1 = R_2 = 10^{-2} \, \mathrm{cm}$ and zero separation gap ($l = 0$), representing the closest-approach scenario.}
    }
    \label{lab_a1_a2_general}
\end{figure*}

{
\subsection{Growth timescales from pebbles to planetesimals}
\label{sec_growth_time}
To assess the efficiency of screening forces in enabling planetesimal formation, we estimate the timescale required for pebbles to grow from centimeter to kilometer scales. This growth occurs through successive binding and merging events: pebbles first form bound pairs with a probability \( \PB \), and subsequently merge via gas drag.
The collisions occur at a rate \( n_{\rm peb} \sigma_{\rm coll} \vrel \), leading to an effective collision timescale of,
\begin{equation}
\tcoll \approx \frac{1}{n_{\rm peb} \sigma_{\rm coll} \vrel}.
\end{equation}
Here we have assumed that $\PB\sim 1$ as is the case for middle disc regions (discussed in next section, Figure \ref{lab_a1_a2_general}). Once a bound state is formed, the pebbles merge on the drag dissipation timescale \( \tdrag \), given by Equation~\ref{eq_tdrag} for considering radius and area of the smaller particle $R=R_2$ and $A=A_2$. The total timescale for a successful growth event through binding and merging is therefore:
\begin{equation}
T_g = \tcoll + \tdrag.
\label{eq_tg}
\end{equation}
}

\section{Results}
\label{sec_results}

We numerically evaluate the binding probability \( \PB \) (Equation~\ref{eq_PB}) to determine the likelihood of pebble coalescence in a protoplanetary disk, incorporating a power-law size distribution of pebbles. For a solar-mass star (\( M_\star = M_\odot \)), we generate a synthetic sample of 100{,}000 pebble pairs with radii \( R_1 \) and \( R_2 \) drawn from this distribution between 0.01 cm and 10 cm. The separation \( l \) between pebbles is randomly selected within the range \( 0 \leq l \leq \lambda(r) \), where \( \lambda(r) \) represents the gas mean free path at radial distance \( r \) in the disk.%
The disk conditions are set with a temperature of \( 200 \, \mathrm{K} \) at \( r = 1 \, \mathrm{AU} \) and a gas surface density of \( \Sigma_0 = 10^{4} \, \mathrm{g\,cm}^{-2} \), which corresponds to a midplane density \( \rho_0 \approx 7 \times 10^{-9} \, \mathrm{g\,cm}^{-3} \) at \( r = 1 \, \mathrm{AU} \) (Equation~\ref{eq_sigma_g}). From observational and empirical constraints, \( \alpha = 10^{-3} \) is a reasonable representative value for turbulence in protoplanetary disks \citep{2011ApJ...727...85H}. We keep pebble density $\ropeb=\,1\mathrm{g\,cm}^{-3} $ throughout the manuscript.
{
Before presenting the general numerical results, we perform a consistency check to assess the physical plausibility of momentum transfer via screening interaction during a close encounter between two pebbles. 
The number of gas-particle scatterings required to significantly alter the pebble's momentum is estimated by the ratio of smaller pebble's momentum attracted by a large pebble (with $M_2<<M_1$) to gas particle momentum:
\[
N_{\rm rec} = \frac{M_2 \, \vrel}{\mu_m m_H v_{\rm th}}.
\]
Here, \( v_{\rm th} = \sqrt{{3 k_B T}/{\mu_m m_H}} \) is the average thermal velocity of gas particles. The actual number of collisions experienced by the smaller pebble due to thermal gas impacts during its interaction with the larger pebble (over a timescale \( (R_1 + \lambda)/\vrel\)) is:
\[
N_{\rm col} = {n \pi R_2^2 v_{\rm th}} \left( \frac{R_1 + \lambda}{\vrel} \right).
\]
Here $n \pi R_2^2 v_{\rm th}$ is the gas particles' collision rate per second on the second pebble's surface. At a representative location in the disk, \( r = 0.5 \, \mathrm{AU} \), we consider a small pebble of radius \( R_2 \sim 0.1 \, \mathrm{cm} \) scattering with a large pebble $R_1=1\, \mathrm{cm} $ with relative velocity \(\vrel \sim 68 \, \mathrm{cm/s} \) (Equation \ref{eq_vrel}), the estimate yields $N_{\rm col}= 6.6 \times 10^{17}$ while  $N_{\rm rec}=3.5 \times 10^{17}$, implying that sufficient scattering events occur to alter the pebble's momentum.
}

\subsection{General scenario}
Figure~\ref{lab_a1_a2_general} illustrates the binding probability \( \PB \) as a function of disk location \( r \) , highlighting the regions where screening forces are most effective. Each point in the figure represents a pebble pair, with the color indicating its binding energy upon merging. {The binding energy (BE) of a pebble pair is computed by setting \( l = 0 \) (i.e., closest contact, \( x = R_1 + R_2 \)) in the right-hand side of Equation~\ref{eq_PE} (approximating $\xmax>>x$):
\be 
\mathrm{BE} = -\frac{\pi n k_B T R_1^2 R_2^2}{R_1 + R_2}
\ee}
For two particles of equal radius, \( R_1 = R_2 = R \), the binding energy scales as \( \mathrm{BE} \propto R^3 \).
{The solid line appearing at the top of the figure corresponds to the binding probability for scattering of largest pebble with $R_1=10 \mathrm{cm} $ with the smallest pebble pair in our sample \( \ie R_2 = 10^{-2} \, \mathrm{cm} \), evaluated under the condition of zero separation (\( l = 0 \)). This configuration represents the most favorable scenario for screening-induced binding, providing an upper bound on the probability across the parameter space.}
Multiple values with a range of \( \PB \) at each radial location \( r \) arise from the randomly selected range of pebble sizes as well as the distances between them, constrained by \( s \leq \lambda \).

{In Figure \ref{lab_a1_a2_general}, we plot the binding probability $\PB$ for for given disk parameters disk surface density \(\Sigma_0\), disk temperature \(T_0\) at 1 AU, and viscosity parameter \(\alpha\). Top left panel has standard values assigned to the disk: \(\Sigma_0 = 10^4 \, \text{g cm}^{-2}\), \(T_0 = 200 \, \text{K}\), and \(\alpha = 10^{-3}\). In the top right panel, we assume a colder disk with \(T_0 = 100 \, \text{K}\). There is no significant difference due to disk temperature because both \(\vcrit\) and \(\vrel\) in Equation \ref{eq_prob_error_func} have identical dependence on disk temperature, thereby canceling the contribution of the disk temperature. The binding probability in the upper panels peaks in the middle disk region (\(r \sim 0.8 \, \mathrm{AU}\), {top panel}), where \(\PB\) approaches close to unity. In the inner disk (\(r < 0.8 \, \mathrm{AU}\)), the probability of binding decreases due to increased fragmentation. This results from the small mean free path (\(\lambda \ll R\)) in the dense inner regions, which reduces the effectiveness of the screening force. {For the assumed disk parameters in this case, no pebble coagulation takes place at \(r < 0.2 \, \mathrm{AU}\).}
Similarly, in the outer disk (\(r > 5 \, \mathrm{AU}\)), the binding probability declines due to the reduced gas pressure in the cold, low-density environment. The middle disk region ($ r= 0.3 $ to a few $ \mathrm{AU}$) is the most favorable zone for pebble binding, where the screening force dominates and enables effective pebble merging. 
{This range is defined as the region where \(\PB\) remains within 90\% of its peak value. Observational findings underscore the importance of the 0.3–10 AU zone in protoplanetary disks, aligning with the commonly referenced ``middle-disk region'' in planet formation studies \citep{2016ApJ...820L..40A, 2018A&A...609L...2B}.}
In the bottom left panel, we reduce the disk density to \(\Sigma_0 = 10^3 \, \text{g cm}^{-2}\) at 1 AU and observe that the peak distribution of the probability shifts to smaller distances \(r \rightarrow 0.2 \, \mathrm{AU}\), while the effective region for pebble coagulation subsequently shifts between 0.08 to 2 AU. }

{The disks may possess greater turbulence, and it is capable of affecting the strength of the screening interaction. In the bottom right panel, we explore the effects of increased turbulence by setting \(\alpha = 10^{-2}\), keeping other parameters the same as in the top left panel. 
While the binding probability is reduced due to higher turbulence causing higher relative velocities between the pebbles (Equation \ref{eq_vrel}), the overall effectiveness of the screening force remains sufficient.}


{Thus, the screening forces continue to operate effectively even in turbulent environments, reinforcing their role as a viable mechanism for pebble aggregation across a wide range of disk conditions.} However, given the long lifetime of the disk (\( \sim \) Myr), even probabilities as low as \( \PB \sim 10^{-2} \) should be considered effective for pebble coalescence, until \( \PB \rightarrow 0 \) at \( r \approx 100 \, \mathrm{AU} \). The binding energies are highest in the inner disk but are associated with low binding probabilities. In contrast, the middle disk region offers moderate binding energies and high probabilities, making it the optimal zone for pebble coagulation.
\subsection{Further growth of larger pebbles to planetesimals}
\label{sec_further_growth}
The results discussed in the previous sections focus on single scattering events, but the formation of planetesimals requires a sustained sequence of interactions that lead to the gradual growth of larger pebbles. From Equation~\ref{eq_vcrit}, we observe that when a small particle of radius \( R_2 \) scatters with a much larger particle, the critical velocity scales as \( v_{\rm crit} \propto 1/\sqrt{R_2} \). This implies that smaller particles can bind more efficiently to larger ones, progressively increasing the probability of further accretion (Equation~\ref{eq_prob_error_func}). Even if kilometer-scale bodies accumulate mass primarily through gravity, their ability to accrete small particles via screening forces remains intact and continues to support their growth.

As a pebble grows, its effective interaction cross-section increases, enhancing its ability to capture additional particles as long as the condition \( s < \lambda \) is satisfied. This sustained accumulation of smaller pebbles onto larger aggregates naturally leads to successive growth, eventually bridging the gap between pebble-sized bodies and kilometer-scale planetesimals. The efficiency of this process depends on the continued availability of smaller particles to maintain a steady supply for further coagulation. {Below, we discuss key aspects including the termination condition for the successive growth and the timescales required conversion of pebbles into planetesimals.}


{ 
\subsubsection{Natural termination of growth via screening and absence of runaway accretion}
The screening mechanism provides an efficient pathway for the aggregation of small particles onto larger bodies in a gas-rich environment. However, the growth process is naturally self-limiting. This section explores the physical conditions under which growth via screening ceases, thereby avoiding unbounded or runaway accretion.
The effectiveness of the screening force relies on the presence of an intervening low-pressure region between two particles, which requires that the slab thickness \( s \) be smaller than the mean free path \( \lambda \) of the surrounding gas. Through successive coagulation with smaller pebbles, when a single pebble grows to a radius \( R_1 \) significantly larger than the surrounding particles (i.e., \( R_1 \gg R_2 \)) and assuming contact interaction (\( l = 0 \)), we have,
\begin{equation}
    s \approx R_1 \left( \sqrt{1 + \frac{2R_2}{R_1}} - 1 \right) \approx R_2.
\end{equation}
Therefore, the requirement for the screening force to operate simplifies to the condition:
\begin{equation}
    R_2 < \lambda.
    \label{eq_r2_term}
\end{equation}
}
{\paragraph{Alternate Derivation of the Condition \( R_2 < \lambda \):}
The same growth termination condition can be independently derived by analyzing the probability \( P_2 \) (Equation~\ref{eq_P_collase}) for avoiding fragmentation during interaction. For \( R_1 \gg R_2 \) and \( R_1 \gg \lmax \), Equation~\ref{eq_P_collase} gives:
\[
P_2 \approx 2 \left( \frac{\lambda - R_2}{R_1} \right).
\]
This expression shows that as \( R_2 \rightarrow \lambda \), the probability \( P_2 \rightarrow 0 \), leading to a vanishing likelihood of successful binding.}

{
This condition thus provides an independent confirmation that only particles smaller than the gas mean free path can be accreted efficiently via the screening force. As the larger body grows and nearby sub-\( \lambda \) particles are consumed, the available population of particles that can be accreted by the larger pebble is depleted, bringing growth to a natural halt.
Consequently, this mechanism does not lead to indefinite or runaway growth. Instead, it supports localized, size-limited accretion events. The availability of sufficiently small particles (\( R_2 < \lambda \)) acts as a bottleneck, ensuring that multiple bodies can grow concurrently in different regions, each terminating growth upon local exhaustion of small projectiles.
This inherent cutoff condition introduces a natural ceiling to growth through screening interactions, distinguishing this process from classical runaway accretion scenarios that typically lack such intrinsic constraints \citep{2000SSRv...92..295W}.
}

{
\subsubsection{Time evolution of pebble size through accretion of smaller pebbles}
\label{sec_time_evo}
}
\begin{figure}[h!]
    \centering
    \includegraphics[width=9 cm, trim = 0 0 0 0, clip]{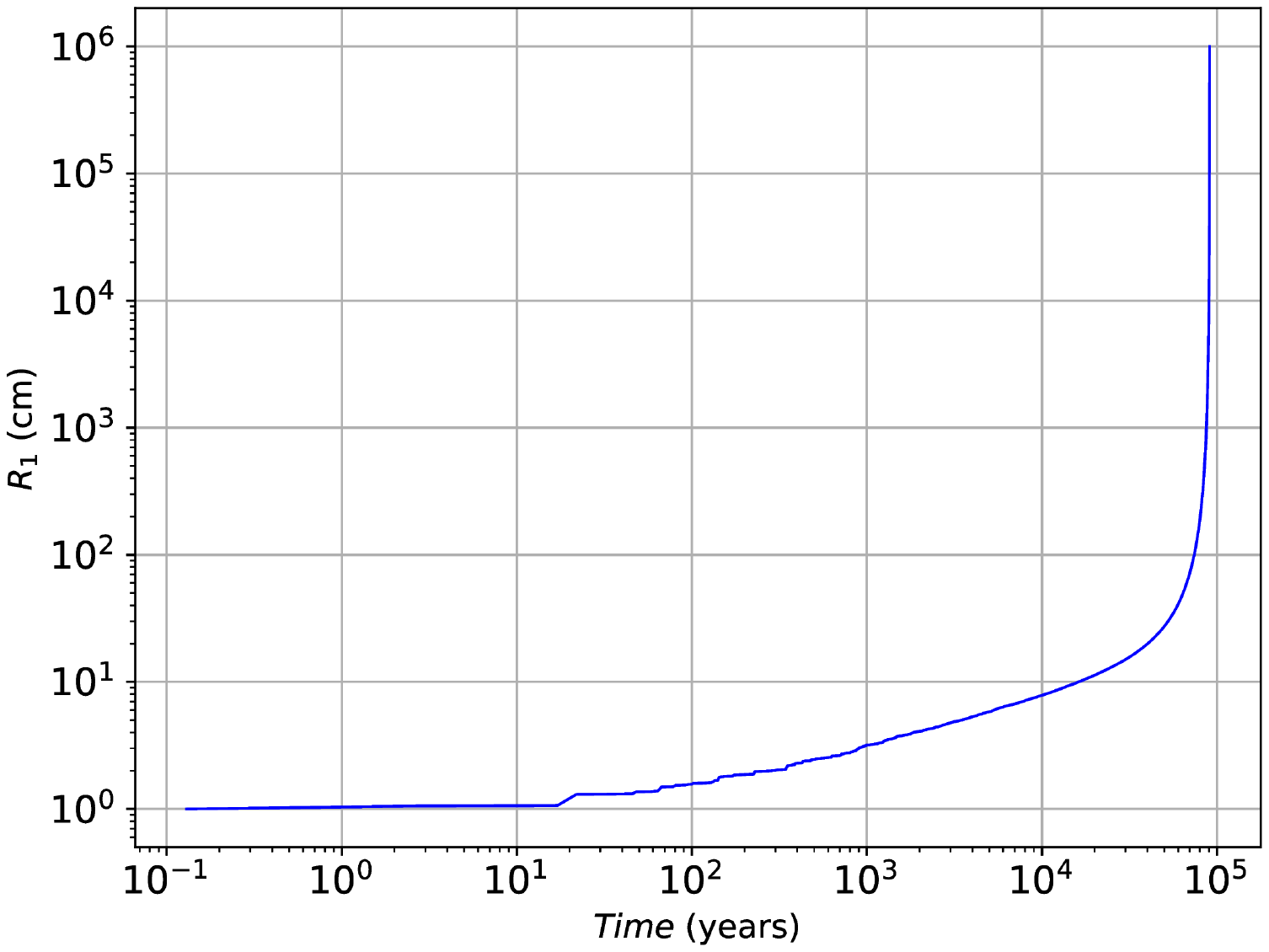}
    \caption{{Growth of a pebble of size $R_1=1$cm to 10 km with time (years) through successive coagulation of particles of sizes $R_2<\lambda$.  The parameters are same as top left panel of Figure \ref{lab_a1_a2_general}.
    }
    }
    \label{lab_size_time_4}
\end{figure}

{To estimate the growth timescale of pebbles in the disk through screening interactions, we consider a pebble of initial size \( R_1 = 1 \, \mathrm{cm} \) at \( r = 1 \, \mathrm{AU} \). This pebble grows through successive mergers with smaller pebbles of size \( R_2 < R_1 \).}
{We plot the growth of the pebble radius \( R_1 \) as a function of time in Figure~\ref{lab_size_time_4}. This growth track is computed numerically by applying Equation~\ref{eq_tg} iteratively with consideration $R_1>>R_2$. After each successful merger, the pebble radius increases according to \( R_1 \rightarrow (R_1^3 + R_2^3)^{1/3} \). The cumulative time is obtained by summing the corresponding values of \( T_g \) over successive growth steps.}

{The pebble requires approximately \( 10^5 \) years to grow from 1 cm to 10 km in size. This timescale is significantly shorter than the typical lifetime of the disk (Myr), indicating that effective growth through screening interactions is achievable. 
As noted above, this growth depends on the continuous availability of particles with sizes \( R_2 < \lambda \) throughout the process.}

\section{Conclusions}
\label{sec_conclusions}

This study investigates a novel mechanism for pebble coagulation in protoplanetary disks, driven by screening forces that arise due to anisotropic gas pressure from mutual shielding of thermal gas flux between pebbles. When the separation between pebbles is smaller than the gas mean free path, screening forces lead to an attractive interaction that facilitates their binding. Once pebbles are bound, gas drag dissipates their relative kinetic energy, leading to rapid merging. This gentle coalescence allows particles to overcome both the fragmentation and bouncing barriers, initiating a cascade of pebble growth that ultimately contributes to planetesimal formation. While weak surface forces, such as van der Waals interactions, may enhance final coalescence, they are not essential for merging, as the screening force is strongest at contact.

We demonstrate that this mechanism is most effective in the middle regions ($ \sim 0.3$ to $10$ AU) of protoplanetary disks, where binding probabilities peak around \(r \sim 0.7 \, \mathrm{AU}\). Screening force emerges naturally from the established physics of gas-rich disks and does not require additional assumptions. Moreover, our results indicate that screening forces remain effective even in highly turbulent environments, with minor reductions in binding probabilities observed at high turbulence levels 
(\(\alpha=10^{-2}\)).

The theoretical predictions of the screening mechanism align well with several observed features of protoplanetary disks, particularly from high-resolution ALMA observations. Observations of disks such as HL Tau, TW Hydrae, HD 163296, and AS 209, where prominent rings, dust enhancements, and pressure bumps have been detected, support the idea that localized, non-gravitational forces contribute to particle concentration. High-resolution ALMA imaging of pressure bumps and dust rings further reinforces the need for additional mechanisms, such as screening forces, that can enhance aggregation in the middle region of the disk \citep{2016ApJ...820L..40A, 2018ApJ...869...17L, 2019ApJ...883...71C}.

The ability of screening forces to facilitate the aggregation of centimeter-sized pebbles is particularly relevant given ALMA’s detection of such grains in disks like HD 163296 and AS 209 through thermal emission and polarization studies \citep{2016ApJ...821L..16C}. These findings support the idea that screening forces can help pebbles overcome key growth barriers and contribute to the formation of planetesimals. {Moreover, the growth process via screening forces does not lead to unbounded, runaway accretion. It naturally terminates once the nearby small particles exceed the local gas mean free path, eliminating further binding opportunities. This size-selective limit ensures parallel growth of multiple aggregates rather than domination by a single body. Notably, our calculations show that particles can grow from centimeter to kilometer sizes within $\sim 10^5$ years, much shorter than typical disk lifetimes, making the process dynamically efficient.
}

The screening forces complement other proposed pathways for planetesimal formation, such as streaming instability and pressure-enhanced vortices. Unlike streaming instability, which requires specific dust-to-gas ratios, screening forces act on individual pebbles, allowing aggregation to occur gradually before large-scale collapse. Additionally,  as the streaming instability enhances local dust density, the screening forces become more effective, further increasing the likelihood of planetesimal formation. 

Replicating the screening interaction in laboratory conditions presents significant challenges due to fundamental differences between protoplanetary disks and laboratory environments. The laboratory setups operate at much higher gas densities, significantly affected by small mean free path. Furthermore, the timescales for relevant interactions in disks span years, whereas laboratory experiments are constrained to much shorter timescales. The small length scales of laboratory environments further complicate the capture of long-range gas-mediated interactions between freely propagating bodies. These challenges explain why screening forces have not yet been observed in experimental settings.

However, their existence can be tested through high-resolution particle-in-cell (PIC) simulations, where macroscopic bodies freely propagate in a gas-rich medium while interacting via resolved gas-particle forces. Such simulations should demonstrate the emergence of an attractive force consistent with our analytical predictions. Future computational studies incorporating detailed gas kinetics and multi-body interactions will be essential for further quantification.
{\paragraph{Caution:} While screening forces provide a viable mechanism for pebble aggregation and are unavoidable natural outcome in the protoplanetary disks, caution is necessary before considering them the sole driver of planetesimal formation. There are several factors beyond the assumptions of this work that can positively or negatively affect the interaction strength:
\begin{enumerate}
\item One potential counteracting effect arises from a fraction of gas particles that elastically scatter off one pebble before colliding with another, potentially introducing a weak repulsive component. However, this effect is expected to be negligible under typical disk conditions, as the fraction of such particles is small compared to the total gas flux contributing to the screening force.


\item In the derivation of the screening force, a radial flux of gas particles is assumed on the surfaces of the spheres. There is a fraction of oblique particle flux that could partially alter the quantitative strength of the force.


\item In this paper, we have described the two-body interaction between pebbles. While the actual interaction may involve simultaneous multi-body interactions, which would enhance the effective force.

\item Shape dependence: The screening forces are evaluated for spherical pebbles. The pebbles have random shapes and the real force might qualitatively differ for different shapes of the pebbles. 

\item The force should have fast but smooth decay at $s\sim \lambda$ such as similar to Yukawa potential $e^{-x/\lambda}$. The precise nature of the force can be explored through high resolution numerical simulations in future. However, the step function assumed at this boundary is reasonable at this stage.
\end{enumerate}
}

{\paragraph{Further:} Future research will focus on testing the efficiency of the screening forces through particle-in-cell (PIC) simulations that incorporate full gas-particle interactions and turbulence effects. {This would test the limiting effects highlighted in the previous paragraph, as well as validate the effectiveness of the screening interaction.}

Beyond its role in planetesimal formation, the screening interaction has broader applications across multiple areas of physics. Since its strength scales with gas pressure, it may be particularly relevant in high-energy astrophysical environments, such as accretion disks around compact objects and high-pressure plasma systems. Additionally, similar effects may arise in terrestrial physics, including soft matter systems, dusty plasmas, and confined particle suspensions in high-pressure gases. Future studies will explore these broader applications in detail.

Additionally, since pebble binding via screening forces extracts thermal energy from the surrounding gas, its impact on local cooling, turbulence suppression, and long-term disk evolution should be explored. Observationally, further multi-wavelength ALMA surveys could help determine whether the predicted pebble concentration effects correlate with observed dust distributions in young planetary systems. In our follow-up papers, we will explore the implications of the screening mechanism in areas beyond planetary sciences.
}

\acknowledgments{{I am grateful to Prof. Asaf Pe'er and Dr. Damien Bègue for valuable discussions, and to Prof. Tapas Das for hosting my visit to the Harish-Chandra Research Institute (HRI), Prayagraj, where part of this work was carried out in the serene environment near the Ganges River.}}

 \bibliography{ref1}{}
 \bibliographystyle{aasjournal}
\end{document}